\documentclass{svproc}
\usepackage{epsfig}
\usepackage{url}

\def\nn{\nonumber}
\newcommand{\be}{\begin{equation}}
\newcommand{\ee}{\end{equation}}
\newcommand{\bea}{\begin{eqnarray}}
\newcommand{\eea}{\end{eqnarray}}
\newcommand{\del}{\partial}
\newcommand{\vk}{\vec {k}}
\newcommand{\om}{\omega}
\newcommand{\ep}{\epsilon}

\begin{document}
\mainmatter              % start of a contribution
\title{Bulk viscosity coefficient of hadronic matter}
\titlerunning{Bulk viscosity of hadronic matter}  % abbreviated title (for running head)
\author{Sabyasachi Ghosh\inst{1,2} \and Sandeep Chatterjee\inst{2}
\and Bedangadas Mohanty\inst{2}}
\authorrunning{S. Ghosh et al.} % abbreviated author list (for running head)
\institute{Department of Physics, University of Calcutta, 92, 
A. P. C. Road, Kolkata - 700009, India\\
%\email{sabyaphy@gmail.com}
\and
School of Physical Sciences, National Institute of Science Education and 
Research Bhubaneswar, HBNI, Jatni, 752050, India}

\maketitle              % typeset the title of the contribution

\begin{abstract}
The bulk viscosity coefficient of hadronic matter has been estimated 
in this present work, where the thermodynamical equilibrium quantity like 
speed of sound in the medium has been obtained by using standard hadron 
resonance gas model. Whereas, the non-equilibrium quantity like thermal
widths of medium constituents have been calculated in the framework field
theory at finite temperature. Our values of bulk viscosity coefficient 
are in agreement with some earlier estimations.

\end{abstract}
\section{Introduction}
Recent research of heavy ion physics has concluded that the medium, formed
in relativistic heavy ion collisions, must have very small shear viscosity,
which is in contrast
to the weak coupling picture, described by the standard finite temperature
calculation of quantum chromo dynamics (QCD). Owing to this motivation,
several microscopic calculations of shear viscosity have been done in recent times.
Estimation of other transport coefficient like bulk viscosity of this
strongly interacting matter is also becoming a contemporary research interest.
In this context, this present article has addressed the bulk viscosity calculation 
for hadronic matter, where
equilibrium thermodynamics for all hadrons in medium are described by Hadron
Resonance Gas (HRG) model. The non-equilibrium part involving the thermal widths of medium
constituents have been calculated by using effective Lagrangian densities 
for the hadronic medium. Here we have assumed pions and nucleons
as the most abundant constituents of the hadronic medium and we have 
calculated their thermal widths,
which basically reflect their in-medium scattering with different mesonic
and baryonic resonances. 
\section{Formalism}
For the equilibrium part of the medium, we have used
the ideal HRG model, where the hadrons and resonances
with masses up-to 2 GeV have been taken from the Particle Data Book.
Constructing total partition function of hadronic medium, one can
easily derived all thermodynamical quantities like pressure ($P$), 
energy density ($\epsilon$), entropy density ($s$). Their temperature
dependence at zero baryon chemical potential ($\mu_B=0$) are quite close
to the corresponding results in the hadronic temperature domain, obtained 
by the lattice quantum chromo dynamics (LQCD) calculations~\cite{LQCD}.
In terms of $P$ and $\ep$, the square of the speed of sound for 
constant baryon density ($n_B$) is defined as
%\be
$v_{n}^2 = \left(\frac{\del P}{\del \epsilon}\right)_{n_B}$~,
%\ee
plays an important role in bulk viscosity calculation.
Either from the Relaxation Time Approximation (RTA) in
kinetic theory approach
or from the one-loop diagram representation in Kubo framework,
one can get standard expressions of bulk viscosity coefficient for pion and nucleon 
components~\cite{Gavin}:
\be
\zeta_\pi = \left(\frac{g_\pi}{T}\right) \int \frac{d^3 \vk}{(2\pi)^3} 
\, \frac{ n_\pi \left[1 + n_\pi\right]}{\om_\pi^2 \, \Gamma_\pi} 
\left\{\left(\frac{1}{3} - v_n^2\right) \vk^2 
- v_n^2 m_\pi^2  \right\}^2
\label{zeta_pi}
\ee

and
\bea
\zeta_N &=& \left(\frac{g_N}{T} \right) \int \frac{d^3\vk}{(2\pi)^3} 
\, \frac{1}{{\om_N^2 \, \Gamma_N}}\left[ 
\left\{ \left(\frac{1}{3} - v_n^2\right) \vk^2 - v_n^2 m_N^2 
- \om_N\left(\frac{\del P}{\del n_B}\right)_{\ep} \right\}^2
n^+_N \left(1 - n^+_N\right) 
\right.\nn\\
&&\left. + \left\{ \left(\frac{1}{3} - v_n^2\right) \vk^2 - v_n^2 m_N^2 
+ \om_N\left(\frac{\del P}{\del n_B}\right)_{\ep} \right\}^2
n^-_N \left(1 - n^-_N\right)\right] ~,
\label{zeta_N}
\eea
where $n_\pi=1/\{e^{\om_\pi/T}-1\}$ with $\om_\pi=\{\vk^2 + m_\pi^2\}^{1/2}$
for pion, $n^{\pm}_N=1/\{e^{(\om_N \mp \mu_B)/T}+1\}$ with $\om_N=\{\vk^2 + m_N^2\}^{1/2}$
for nucleon and their respective degeneracy factors are $g_\pi=3$ and $g_N=2\times 2$.
The $m_\pi$, $m_N$ stand for masses of pion, nucleon and their momenta are denoted by $\vk$.
The $\Gamma_\pi$ and $\Gamma_N$ are thermal widths of pion and nucleon, which have
been calculated from the imaginary part of their self-energies (on-shell)
at finite temperature. We have considered different
mesonic and baryonic loops for pion self-energy, whereas, pion-baryon intermediate states
are taken in nucleon self-energy~\cite{G_N,G_NISER}.

\section{Results and Conclusion}
\begin{figure}
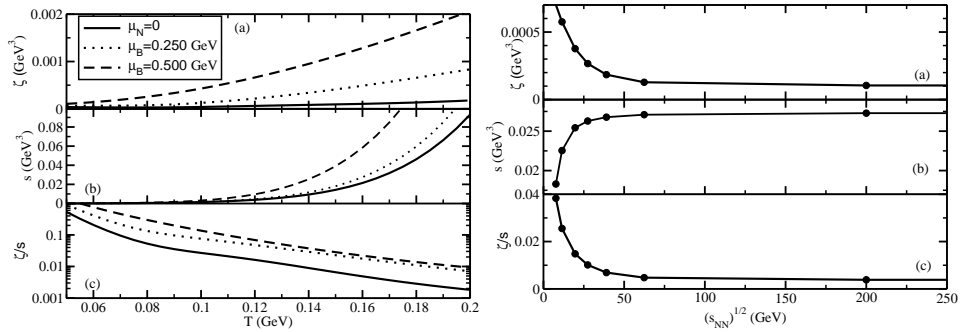

%\begin{center}
\includegraphics[scale=0.25]{z_s_Tmu.eps}
\includegraphics[scale=0.25]{DAEz_s.eps}
\caption{Left : $T$ dependence of bulk viscosity (a), entropy density (b) and their 
ratio $\zeta/s$ (c) at $\mu_B=0$ (solid line), $0.250$ GeV 
(dotted line) and $0.500$ GeV (dash line).
Right : Variation of same quantities $\zeta$ (a), $s$ (b) 
and $\zeta/s$ (c) with the variation of center of mass energy ($\sqrt{s_{NN}}$).} 
\label{z_s_Tmu}
%\end{center}
\end{figure}
Using the explicit momentum $\vk$, $T$ and $\mu_B$ dependent thermal
widths of pion and nucleon in the integrands of Eqs.~(\ref{zeta_pi})
and (\ref{zeta_N}), we can calculate bulk viscosity of pion and nucleon components
respectively. 
%
%These thermal widths, which signify the in-medium probabilities 
%of pion and nucleon scatterings with different mesonic and baryonic resonances
%of the medium, inversely control the numerical strengths of bulk viscosities.
%
Adding these two components, we get total bulk viscosity, which is plotted
in the left panel of Fig.~\ref{z_s_Tmu}(a). The entropy density $s$, obtained
from HRG and the ratio $\zeta/s$ are also drawn in the left panel of 
Fig.~\ref{z_s_Tmu}(b) and (c), which reveal their temperature dependence at 
$\mu_B=0$ (solid line), $0.250$ (dotted line) and $0.500$ GeV (dashed line).
We notice that $\zeta$ and $s$ both increase with $T$ and $\mu_B$.
Whereas, $\zeta/s$ is a decreasing function of $T$ (but still 
increase in $\mu_B$) because the increment of $s(T)$ is larger than 
the increment of $\zeta(T)$.

Next, right panel of Fig.~\ref{z_s_Tmu}(a), (b) and (c) show respectively
the variation of $\zeta$, $s$ and $\zeta/s$ with the variation of 
center of mass energy $\sqrt{s_{NN}}$
%
%\footnote{
%Reader are
%requested to be careful on the same symbol $s$, used for entropy density and square of
%beam energy.
%}.
%
We have used the parameterization from Ref.~\cite{HRGKarsch},
where beam energy dependence of $T$ and $\mu_B$ used in computation 
are those obtained from fits to hadron yields. 
We notice in the right panel of Fig.~\ref{z_s_Tmu} that $\zeta$ (a) as well as $\zeta/s$ (c) 
are decreasing with $\sqrt{s_{NN}}$, which is qualitatively agreeing with the 
results of earlier Ref.~\cite{Kadam2}. 
The decreasing trend of $\zeta$ and $\zeta/s$ with $\sqrt{s}$ can 
be understood from the fact that $\mu_B$ decreases with $\sqrt{s}$ while $T$ remains 
fairly constant in the range of $\sqrt{s}$ analyzed here and the $\zeta$ and 
$\zeta/s$ decrease with decreasing of $\mu_B$ as we have noticed in the left panel
of Fig.~(\ref{z_s_Tmu}). At $\mu_B=0$, the decreasing nature of our $\zeta/s(T)$
agrees with most of the earlier works as addressed elaborately in Ref.~\cite{G_NISER}.
However, some investigations have reported it to increase with $T$ and 
some time also have a peak structure
near transition temperature~\cite{Kinkar}.

{\bf Acknowledgments:} Authors thank UGC, DAE, DST Govt. of India for financial support
during the period, when this work was carried out.



\begin{thebibliography}{6}
%
\bibitem{LQCD}Bazavov, A. et al. (HotQCD Collaboration) :
Equation of state in (2+1)-flavor QCD.
Phys. Rev. {\bf D 90}, 094503 (2014).
%
\bibitem{Gavin}Gavin, S. : Transport coefficients in ultra-relativistic
heavy-ion collisions.
Nucl.Phys. {\bf A 435} (1985) 826.
%
\bibitem{G_N}Ghosh, S. :
The nucleon thermal width due to pion-baryon loops and its contributions in Shear viscosity.
Phys. Rev. {\bf C 90}, 025202 (2014).
%
\bibitem{G_NISER}Ghosh, S., Chatterjee, S., Mohanty, B. :
Bulk viscosity for pion and nucleon thermal fluctuation in the hadron resonance gas model.
Phys. Rev. {\bf C 94}, 045208 (2016).
%
\bibitem{HRGKarsch}Karsch, F. and Redlich, K. :
Probing freeze-out conditions in heavy ion collisions
with moments of charge fluctuations.
Phys. Lett. {\bf B 695}, 136 (2011).
%
\bibitem{Kadam2}Kadam, G. P., Mishra, H. :
Dissipative properties of hot and dense hadronic matter in an excluded-volume
hadron resonance gas model.
Phys. Rev. {\bf C 92} (2015) 035203.
%
\bibitem{Kinkar}Saha, K., Upadhaya, S., Ghosh, S. :
A comparative study on two different approaches of bulk viscosity
in the Polyakov Nambu Jona-Lasinio model.
Mod. Phys. Lett. {\bf A 32}, 1750018 (2017).
%
%
%\bibitem {smit:wat}
%Smith, T.F., Waterman, M.S.: Identification of common molecular subsequences.
%J. Mol. Biol. 147, 195?197 (1981). \url{doi:10.1016/0022-2836(81)90087-5}
%
%\bibitem {may:ehr:stein}
%May, P., Ehrlich, H.-C., Steinke, T.: ZIB structure prediction pipeline:
%composing a complex biological workflow through web services.
%In: Nagel, W.E., Walter, W.V., Lehner, W. (eds.) Euro-Par 2006.
%LNCS, vol. 4128, pp. 1148?1158. Springer, Heidelberg (2006).
%\url{doi:10.1007/11823285_121}
%
%\bibitem {fost:kes}
%Foster, I., Kesselman, C.: The Grid: Blueprint for a New Computing Infrastructure.
%Morgan Kaufmann, San Francisco (1999)
%
%\bibitem {czaj:fitz}
%Czajkowski, K., Fitzgerald, S., Foster, I., Kesselman, C.: Grid information services
%for distributed resource sharing. In: 10th IEEE International Symposium
%on High Performance Distributed Computing, pp. 181?184. IEEE Press, New York (2001).
%\url{doi: 10.1109/HPDC.2001.945188}
%
%\bibitem {fo:kes:nic:tue}
%Foster, I., Kesselman, C., Nick, J., Tuecke, S.: The physiology of the grid: an open grid services architecture for distributed systems integration. Technical report, Global Grid
%Forum (2002)
%
%\bibitem {onlyurl}
%National Center for Biotechnology Information. \url{http://www.ncbi.nlm.nih.gov}
%

\end{thebibliography}
\end{document}